\documentclass[
    ,final            % use final for the camera ready runs
  ]
  {aipproc}

\layoutstyle{8x11double}

\begin{document}

\title{CTA simulations with CORSIKA/sim\_telarray}

%\classification{<Replace this text with PACS numbers; choose from this list:
%                \texttt{http://www.aip..org/pacs/index.html}>}

\author{K. Bernl{\"o}hr}{
  address={Max-Planck-Institut f\"ur Kernphysik, Postfach 103980, 69029 Heidelberg, Germany}
  , altaddress={Institut f\"ur Physik, Humboldt-Universit\"at zu Berlin, Newtonstra{\ss}e 15, 12489 Berlin, Germany}
  % , email={Konrad.Bernloehr@mpi-hd.mpg.de}
%\author{K. Bernl{\"o}hr$^{*\dagger}$}{
%  address={$^{*}$ Max-Planck-Institut f\"ur Kernphysik, Postfach 103980, 69029 Heidelberg, Germany}
%  , altaddress={$^{\dagger}$ Institut f\"ur Physik, Humboldt-Universit\"at zu Berlin, Newtonstra{\ss}e 15, 12489 Berlin, Germany}
%  % , email={Konrad.Bernloehr@mpi-hd.mpg.de}
}

%%%%%%%%%%%%%%%%%%%%%%%%%%%%%%%%%%%%%%%%%%%%%%%%%%%%%%%%%%%%%%%%%%%%%%%%%

\begin{abstract}
While current atmospheric Cherenkov installations consist of only
a few telescopes each, future installations will be far more complex.
Monte Carlo simulations have become an essential tool for the
design and optimisation of such installations.
The CORSIKA air{}-shower simulation code and the {\tt sim\_telarray} code for
simulation of arrays of Cherenkov telescopes have been used to simulate
several candidate configurations of the future Cherenkov Telescope
Array (CTA) in detail. Together with other detailed and simplified
simulations the resulting data provide the basis for the ongoing
optimisation of CTA over a wide energy range. In this paper, the
simulation methods are outlined and preliminary results on a number
of configurations are presented. It is demonstrated that the
initial goals of the CTA project can be achieved with available
technology, at least in the medium and high energy range (about 100 GeV to
100 TeV).
\end{abstract}

\maketitle

%%%%%%%%%%%%%%%%%%%%%%%%%%%%%%%%%%%%%%%%%%%%
%% MAINMATTER
%%%%%%%%%%%%%%%%%%%%%%%%%%%%%%%%%%%%%%%%%%%%

%%%%%%%%%%%%%%%%%%%%%%%%%%%%%%%%%%%%%%%%%%%%%%%%%%%%%%%%%%%%%%%%%%%%%%%%%

\section{Introduction}

The Cherenkov Telescope Array (CTA) project design study 
\cite{cta-url} is under way to
design and optimise the next generation of stereoscopic Imaging
Atmospheric Cherenkov Telescope (IACT) arrays. CTA aims to achieve
an order of magnitude improvement in point-source sensitivity 
over current instruments and a
wider energy coverage in order to study astrophysical processes of
more sources in more detail.

While extrapolation from current generation instruments may be a first
guide to what is needed to achieve the CTA goals, only detailed
simulations can give definite answers. Physical processes in air
showers may ultimately limit the gamma{}-hadron separation power of
the instruments, as well as their angular and energy resolution.
Depending on area coverage and instrumental capabilities, any
affordable solution will typically not reach such ultimate limits.
The approach in this paper is the evaluation of realistic -- or conservative --
performance parameters by detailed simulations of air showers and
the potential instruments, followed by analysis methods well tested
on current instruments.

The air{}-shower simulation is based on the CORSIKA \cite{CORSIKA} code with its
choice of different interaction models. With the IACT option of CORSIKA
we can simulate the Cherenkov light hitting arbitrary arrays of
telescopes, each defined separately by its position and a fiducial radius. 
Since the CORSIKA stage requires the largest amount of
CPU time, each shower is re-used at different random displacements.
Due to the excellent gamma-hadron separation of all the
configurations envisaged, a huge number of
hadron showers has to be simulated in order to estimate the
remaining backgrounds -- in most cases billions of events.

\begin{figure}
\includegraphics[width=\hsize]{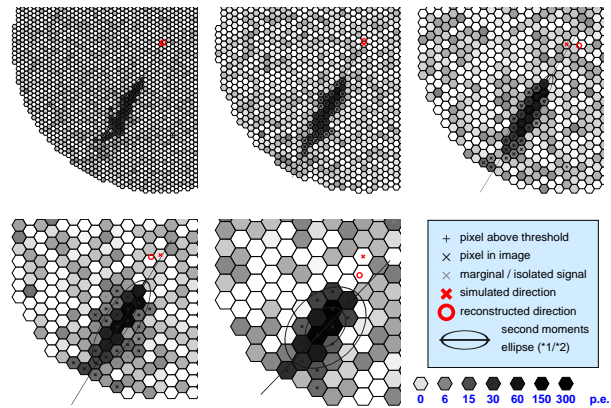}
\caption{Different pixel sizes can have an
impact on the image obtained from the same shower (here
0.07{\textdegree} to 0.28{\textdegree} pixels on a 23 m telescope, only
part of f.o.v. shown). CORSIKA output can be piped directly into
multiple telescope simulations running in parallel, allowing for
efficient simulation of different configurations. 
(See Figure \ref{fig:pixel-size-results} for results.)
\label{fig:pixel-size-image}}
\end{figure}

In a second processing stage, the atmospheric extinction
and the details of the detector response are simulated by the
{\tt sim\_telarray} program \cite{simtelarray}. 
This program is very flexible and
different detectors with different read{}-out, different triggering
schemes, or just different reflectors are treated by the same code,
by just specifying different configuration files. Simulations include
optical ray{}-tracing, night{}-sky background, all relevant electronic
pulse shapes, switching behaviour of comparators or discriminators, as
well as many other details. Telescope configurations are tested
for random night-sky trigger rates before being used for shower
simulations.

\begin{figure}
\includegraphics[width=0.7\textwidth]{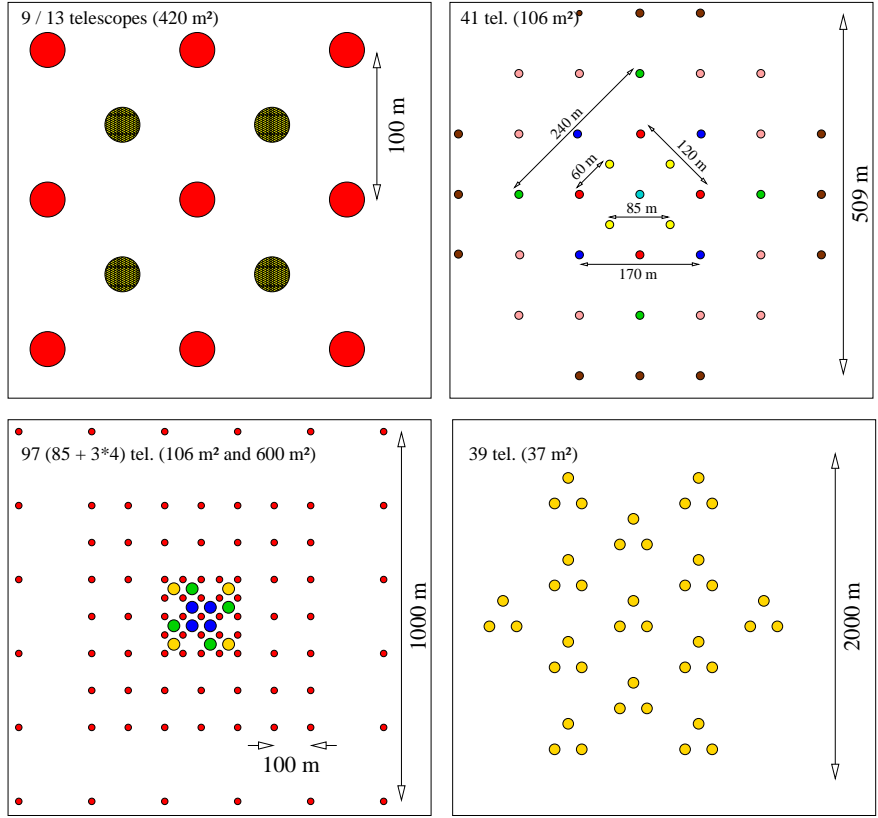}
\caption{Selection of different array
configurations simulated for altitudes of 1800 to 2000 m.
See text for more details.
\label{fig:arrays}}
\end{figure}

For efficiency reasons, the CORSIKA output data is usually piped into the
telescope simulation, without intermediate on-disk storage.
By placing a utility program termed {\tt multipipe\_corsika} into
this pipe, the CORSIKA data can be piped into multiple telescope
simulations at the same time, e.g. for different pixel sizes (see
Figure \ref{fig:pixel-size-image}),
different offset angles etc.

%%%%%%%%%%%%%%%%%%%%%%%%%%%%%%%%%%%%%%%%%%%%%%%%%%%%%%%%%%%%%%%%%%%%%%%%%

\section{A variety of candidate configurations}

Since the performance is a complex function of array layout (which may
consist of multiple types of telescopes), and of a variety of
telescope parameters, detailed simulations can be performed only
for a limited {--} and therefore rather complementary {--} subset of layout and
telescope parameters. These candidate configurations 
(see Figure \ref{fig:arrays}) include:
\begin{itemize}
\item A 9{}-telescope array aiming at low energies (420 m{\texttwosuperior}
mirror area each, with 0.1{\textdegree} pixels covering a
5{\textdegree} field{}-of{}-view (f.o.v.). This configuration was
simulated at different altitudes (2000 m, 3500 m, and 5000 m
a.s.l.). The 2000 m simulations were carried out with a variety of
pixel sizes (from 0.07{\textdegree} up to 0.28{\textdegree}, see 
Figure~\ref{fig:pixel-size-image}). It was also cross-checked with
an different telescope simulation code \cite{ICRC2007paper}.
\item 
An array of 41 H.E.S.S.{}-I type telescopes (106 m{\texttwosuperior}),
to see how gamma{}-hadron rejection can be improved by better coverage
with current{}-generation instruments. It also allows to test
arrays of 4, 5, 9, or 16 telescopes with different inter-telescope
separations by selecting subset of the 41 telescopes in the
analysis.
\item 
An array consisting of telescopes of two different sizes (600
m{\texttwosuperior} and 106 m{\texttwosuperior} mirror area,
5{\textdegree} and 7{\textdegree} f.o.v.) aiming at a rather wide
energy coverage. For the large telescopes, three sets of four
telescopes each, with different separations, were included in
the simulation -- usually selecting only one set of four large
plus the 85 small telescopes in the analysis.
All telescopes were assumed to have a 50\% higher quantum efficiency
than H.E.S.S.
\item 
Different arrays of rather small (37 m{\texttwosuperior}) telescopes
with large separations and wide{}-f.o.v. cameras (up to 8{\textdegree}
with 0.3{\textdegree} pixels), aiming at high energies. 
\end{itemize}

Simulations include gammas, protons, and electrons as
primaries {--} nuclei, being generally easy to distinguish from
gammas, were not considered. 
All simulations were carried out at the Max Planck Institute in
Heidelberg, except for the 3500 m 9{}-telescope configuration which was
processed at SLAC.

%%%%%%%%%%%%%%%%%%%%%%%%%%%%%%%%%%%%%%%%%%%%%%%%%%%%%%%%%%%%%%%%%%%%%%%%%

\section{Analysis}

The analysis of simulation data is based on methods similar to the
H.E.S.S. Hillas{}-parameter based standard analysis, with some
extensions like using the height of shower maximum and how well the
lateral distribution conforms with expectations for gamma showers.
Image shape cuts include the mean reduced scaled width and length
\cite{HESS-Crab-paper}. Since, after geometrical shower reconstruction,
the image in each telescope can be used for an energy estimate,
the consistency of the individual estimates for the same event 
is used as another cut. Showers where the expected accuracy of
the final (weighted mean) energy estimate is poor for the
given energy, e.g. showers outside the array, are also discarded.
At the lowest energies, a few tens of GeV, the image shape cuts
typically have little gamma-hadron discrimination power. 
The most important parameter in this energy range turned out to
be the height of shower maximum.
Selection cuts were manually optimised as simple functions of
reconstructed energy or telescope multiplicity (e.g. of the
form $a+b\;\log E$, with additional lower and upper bounds).
Further improvements
by more sophisticated shower reconstruction and advanced cut
optimisations may be feasible. In that sense, the following
performance results are very conservative.

The set of cuts from the H.E.S.S. standard analysis was also
used on H.E.S.S.-I simulations as well as corresponding
four-telescope subsets of the 41-telescope array, with 
corresponding sensitivities matching the actual H.E.S.S.
sensitivity \cite{HESS-Crab-paper}.

\begin{figure}
\includegraphics[width=\hsize]{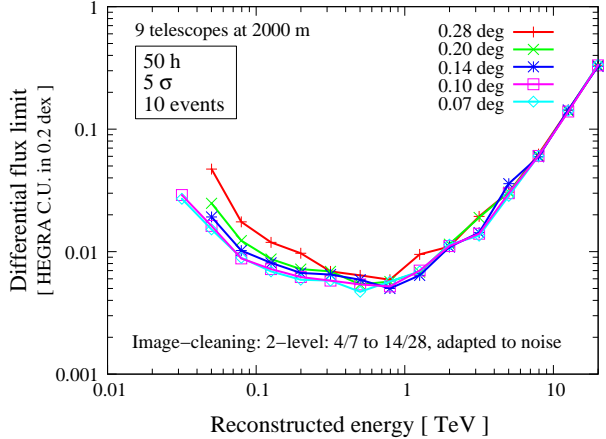}
\caption{Differential sensitivity limit
(with 5 sigma, 10 events in each energy bin) in HEGRA Crab units 
\cite{HEGRA-Crab} for
the 9{}-telescope array at 2000 m altitude, at a zenith
angle of 20{\textdegree}. Each curve is for one of
the pixel sizes in Figure \ref{fig:pixel-size-image}. Since the
larger pixels will see more night-sky background noise, it is
important to adapt image cleaning thresholds to this noise.
Only at very low energies any improvement
could be achieved by small pixels.
\label{fig:pixel-size-results}}
\end{figure}

%%%%%%%%%%%%%%%%%%%%%%%%%%%%%%%%%%%%%%%%%%%%%%%%%%%%%%%%%%%%%%%%%%%%%%%%%

\section {Performance results}

The performance of the tested CTA candidate configurations was
evaluated for point sources, usually taking advantage of the angular
resolution improving with increasing number of telescopes with usable
images. Since the rejection of hadronic background improves with
increasing energy, most selection cuts have to be rather strict at
low energies and get looser towards higher energies. Both proton and
electron background was included, with the electron flux
extrapolated ${\propto}{E}^{-3.3}$ beyond available measurements. 
In contrast to na\"ive expectations, electrons never dominated
the background at the lowest energies, due to the energy-dependent
hadron rejection efficiency.
In the evaluation of flux limits, the
Li\&Ma formula was used to find the minimum number of gammas to achieve
a 5{}-sigma significance. In addition, a minimum of 10 gammas was required. At
very low energies, another limitation had to be taken into account: the
systematics in the subtraction of remaining background, here
assumed at a level of 1\% being equivalent to a 1{}-sigma fluctuation.

It turns out that with the arrays of nine 420~m{\texttwosuperior} telescopes
the energy threshold can in fact be reduced to some 20~GeV. Rather strict
selection cuts are needed at the lowest energies to achieve
sufficient gamma-hadron separation to avoid running into the
limitation by background systematics. 

The 41-telescope array of 
106~m{\texttwosuperior} telescopes would not substantially improve the 
energy threshold compared to the four-telescope H.E.S.S. phase I array.
Thanks to the much better sampling of the showers it would however
improve gamma-hadron separation, angular resolution, and effective
area compared to H.E.S.S.-I, resulting in a sensitivity improvement
much faster than the na\"ive $1/\sqrt{N}$ expectation ($N$ being
the number of telescopes).

The 97-telescope array (of which usually 85 small plus 4 large
telescopes were used in the analysis) combines low energy threshold
with much improved gamma-hadron rejection. In fact, the presence
of the smaller telescopes helps to improve the gamma-hadron
rejection even in the energy domain normally only accessible
to the large telescopes -- by rejecting hadron showers where
a sub-shower may look like a low-energy gamma-ray shower.
At intermediate energies, this configuration fully meets
initial goals for CTA (`` milli-Crab sensitivity'').

At the highest energies envisaged for CTA, even the square kilometre
area of the 97-telescope configuration may not be sufficient to
detect enough events. It could be supplemented with additional
small telescopes instrumented with much coarser pixels (up to 
0.3{\textdegree}) and operated at larger inter-telescope separations.
This is demonstrated by the assumed 39-telescope array.

\begin{figure}
\includegraphics[width=\hsize]{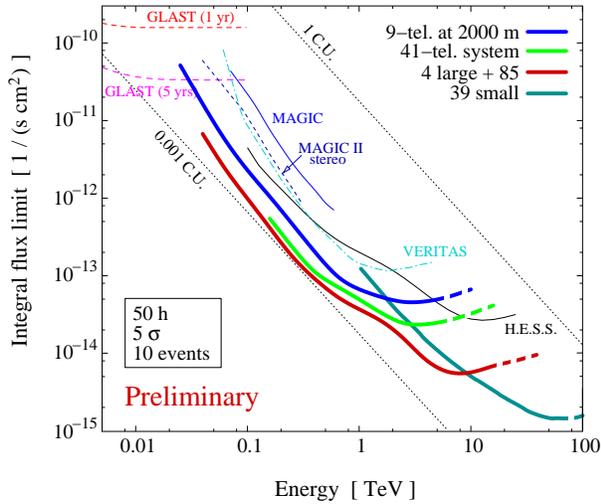}
\caption{Point source sensitivity limits
F({\textgreater}E) of the four arrays from Figure \ref{fig:arrays} 
at 20{\textdegree} zenith angle, 
compared with current instruments 
\cite{HESS-Crab-paper,MAGIC-II-sens,VERITAS-proposal,GLAST-sens}. 
One C.U. (Crab unit) in this case is
$F_{\rm{cu}}(>E)=1.78\cdot10^{-7}(E/\mathrm{TeV})^{-1.57}\mathrm{photons}/(\mathrm{m}^{2}\;\mathrm{s})$
\cite{HEGRA-Crab}.
\label{fig:sensitivity-comparison}}
\end{figure}

\begin{figure}
\includegraphics[width=\hsize]{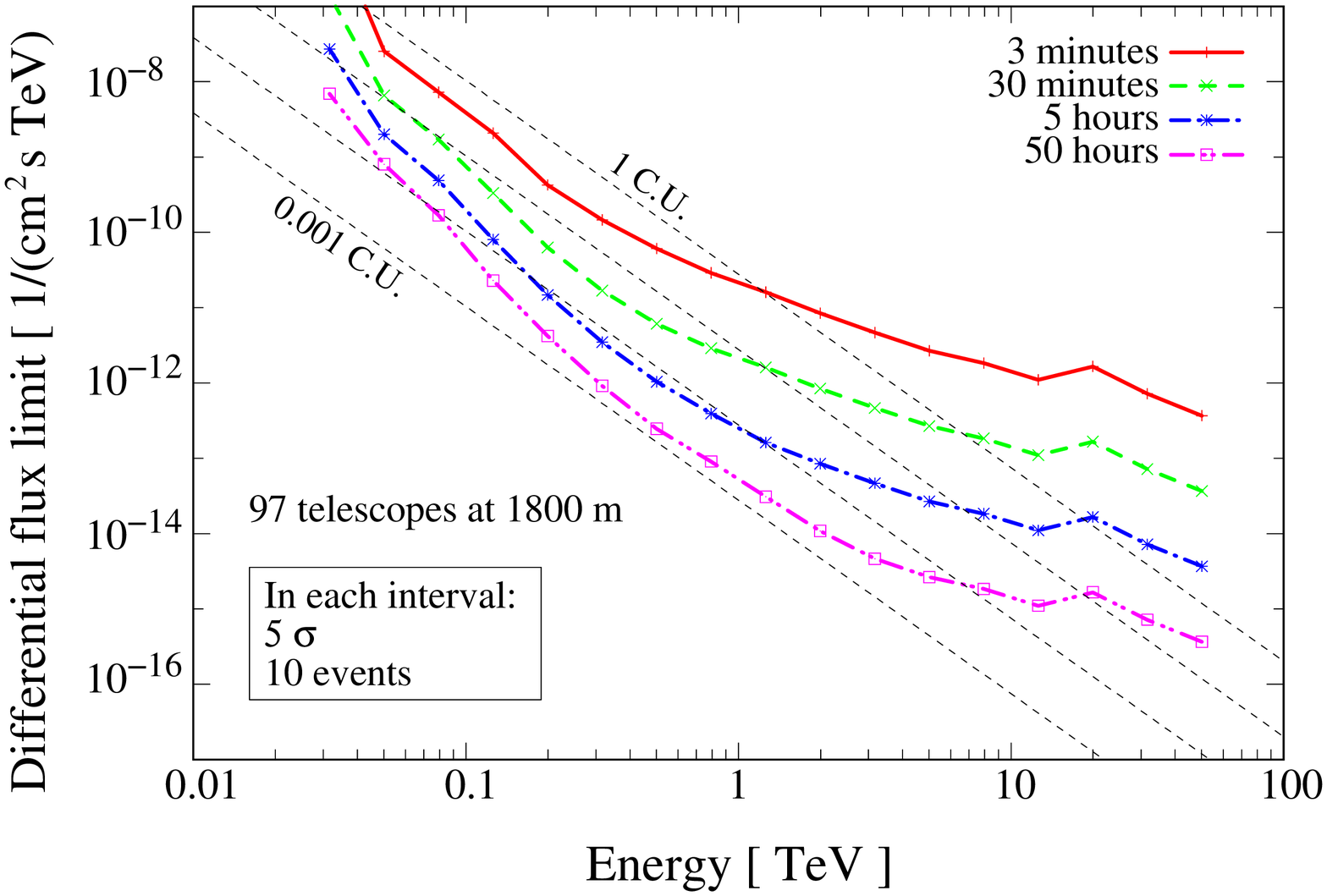}
\caption{Differential sensitivity limit
(with 5 sigma, 10 events in each energy bin of 0.2 dex) for the 97{}-tel.
configuration, with different exposure times, from 3 minutes to 50 hours. 
A 1 hour exposure would
be enough for a high{}-quality spectrum of a 0.1 Crab source over two
orders of magnitude in energy. 50 hours would be needed for the
spectrum of a milli{}-Crab source.
One C.U. (Crab unit) is assumed here as 
$dF_{\rm{cu}}/dE=2.79\cdot10^{-7}(E/\mathrm{TeV})^{-2.57}\mathrm{photons}/(\mathrm{m}^{2}\;\mathrm{s}\;\mathrm{TeV})$
\cite{HEGRA-Crab}.
\label{fig:sensitivity-time}}
\end{figure}

%%%%%%%%%%%%%%%%%%%%%%%%%%%%%%%%%%%%%%%%%%%%%%%%%%%%%%%%%%%%%%%%%%%%%%%%%

\section {Conclusions and outlook}

The combination of the CORSIKA and sim\_telarray programs is very well
suited for simulations of large telescope arrays, even with hundreds
of telescopes. It is also very
flexible and can be adapted to arbitrary array configurations and
telescope setups just by means of run{}-time configuration. The
subsequent analysis with its Hillas{}-parameter based shower
reconstruction results in rather conservative performance
estimates. Arrays consisting of multiple telescope types are easily
handled in all stages. First simulations included a range of
different test configurations, with emphasis on different energy
ranges, and demonstrated that the CTA sensitivity goal can be
achieved \cite{ICRC2007paper} {--} at least for energies above 50 to 100 GeV. Future
simulations will include more realistic CTA configurations (within
budget constraints) as well as some corner cases needed to improve the
CTA design optimisation scheme.

%%%%%%%%%%%%%%%%%%%%%%%%%%%%%%%%%%%%%%%%%%%%%%%%%%%%%%%%%%%%%%%%%%%%%%%%%

%%%%%%%%%%%%%%%%%%%%%%%%%%%%%%%%%%%%%%%%%%%%%%%%
%% BACKMATTER
%%%%%%%%%%%%%%%%%%%%%%%%%%%%%%%%%%%%%%%%%%%%%%%%

\begin{theacknowledgments}
I would like to thank Emiliano Carmona, Jim Hinton, and many others
for useful discussions on the CTA configurations. Stefan Funk has
complemented my 9-telescope simulations for 2000 and 5000~m altitude
with corresonding ones for 3500~m altitude, carried out at the
SLAC computing center.
\end{theacknowledgments}

%%%%%%%%%%%%%%%%%%%%%%%%%%%%%%%%%%%%%%%%%%%%%%%%%%%%%%%%%%%%%%%%%%%%%%%%%


\begin{thebibliography}{9}

\bibitem{cta-url}
\url{http://www.cta-observatory.org/}

\bibitem{CORSIKA}
D. Heck et al., ``CORSIKA: A Monte Carlo code to simulate extensive air showers''.
Technical Report FZKA 6019, Forschungszentrum Karlsruhe, 1998.

\bibitem{simtelarray}
K. Bernl\"ohr, Astroparticle Physics (2008).
doi:10.1016/j.astropartphys.2008.07.009 
[arXiv:0808.2253]

\bibitem{ICRC2007paper}
K. Bernl\"ohr, E.~Carmona et al.,
``MC simulations and layout studies for a future Cherenkov Telescope Array'',
in \emph{Proc. of the 30$^{\it th}$ International Cosmic Ray Conference},
M\'erida, Mexico, 2007.

\bibitem{HEGRA-Crab}
F.A.~Aharonian et~al. {(HEGRA Collaboration)}.
\newblock {\em Astroph.J.} {\bf 539}, 317--324 (2000).

\bibitem{HESS-Crab-paper}
F.~Aharonian et~al. {(H.E.S.S. Collaboration)}.
\newblock {\em Astron. Astroph.} {\bf 457}, 899--915 (2006).

\bibitem{MAGIC-II-sens}
C.~Baixeras et~al,
\newblock ``{MAGIC} phase {II}'',
\newblock in {\em Proc. of the 29$^{\it th}$ International
Cosmic Ray Conference}, {P}une, India, 2005, volume~5, pp. 227--230.

\bibitem{VERITAS-proposal}
T.C. Weekes et~al.
\newblock ``{VERITAS} proposal''.
\newblock Technical report, 2000.

\bibitem{GLAST-sens}
Toby Burnett.
\newblock Technical Report LAT DOC AM-04369, University of Washington, 2004.

\end{thebibliography}
\end{document}